\documentclass[12pt,english,british]{iopart}
\usepackage{babel}
\usepackage{lmodern}
\usepackage{textcomp}
\usepackage{graphicx}
\usepackage{iopams}
\usepackage{cite}
\usepackage{subfigure}
\usepackage{url}
\usepackage{epstopdf}

\newcommand{\E}[1]{\cdot10^{#1}}
\newcommand{\too}{\longrightarrow}

\begin{document}

\title[Time-resolved OES of nanosecond pulsed discharges in N$_{2}$ and N$_{2}$/H$_{2}$O mixtures]{Time-resolved optical emission spectroscopy of nanosecond pulsed discharges in atmospheric pressure N$_{2}$ and N$_{2}$/H$_{2}$O mixtures}

\author{R.M. van der Horst, T. Verreycken, E.M. van Veldhuizen and P.J. Bruggeman}

\address{Eindhoven University of Technology, Department of Applied Physics, PO Box 513, 5600 MB Eindhoven, The Netherlands}

\ead{r.m.v.d.horst@tue.nl}

\begin{abstract}
In this contribution, nanosecond pulsed discharges in N$_{2}$ and N$_{2}$/0.9\% H$_{2}$O at atmospheric pressure (at 300~K) are studied with time-resolved imaging, optical emission spectroscopy and Rayleigh
scattering. A 170~ns high voltage pulse is applied across two pin-shaped electrodes at a frequency of 1~kHz. The discharge consists of three phases: an ignition phase, a spark phase and a recombination phase. During the ignition phase the emission is mainly caused by molecular nitrogen (N$_{2}$(C-B)). In the spark and recombination phase mainly atomic nitrogen emission is observed. The emission when H$_{2}$O is added is very similar, except the small contribution of H$_{\alpha}$ and the intensity of the molecular N$_{2}$(C-B) emission is less.

The gas temperature during the ignition phase is about 350~K, during the discharge the gas temperature increases and is 1~\textmu{}s after ignition equal to 750~K. The electron density is obtained by the
broadening of the N emission line at 746~nm and, if water is added, the H$_{\alpha}$ line. The electron density reaches densities up to $4\E{24}$~m$^{-3}$. Addition of water has no significant influence on the gas temperature and electron density.

The diagnostics used in this study are described in detail and the validity of different techniques is compared with previously reported results of other groups.
\end{abstract}

\maketitle

\section{Introduction}

Nanosecond repetitively pulsed (NRP) discharges are of increasing interest in a broad range of biomedical~\cite{Ayan2008}, combustion~\cite{Adamovich2009} and environmental applications~\cite{Pemen2003}. NRP discharges are created by repetitively applying a short high voltage pulse across two electrodes. The high electrical field causes ionization and dissociation processes. In spite the fact that the applied electrical field is short and often rather moderate compared to pulsed coronas for which tens of kV are applied, the pre-ionization by the previous discharge helps to make the ignition of NRP discharges reproducible. The short voltage duration prevents the transition to an arc and excessive heating~\cite{Bruggeman2009}. These discharges are also a rich source of reactive radicals~\cite{Stancu2010}. One of the possible applications is air cleaning. Pollutants can be removed from the air due to reactions with the radicals created by the discharge~\cite{Winands2006}. NRP discharges are also studied extensively in the field of plasma enhanced combustion~\cite{Adamovich2009,Laux2003,Pai2010}.
Due to the presence of radicals, the efficiency of the combustion can be increased.

Various properties of NRP discharges have already been reported in literature. The electron density in a NRP discharge (spark regime) in preheated air at atmospheric pressure has been studied by e.g.\ Pai~\etal\cite{Pai2010} with space- and time-resolved OES. They produced the discharge by applying short voltage pulses (7-8~kV) of 10~ns with a frequency of 30~kHz across two pin electrodes. They found that the spark generates a spatial homogeneous emission of N$_{2}$(C-B) and N$_{2}^{+}$(B-X). Further it was demonstrated that a spark can form without a streamer if repetitive pulsing is used. They reported electron densities, obtained from the conduction current, between $10^{21}$ and $10^{22}$~m$^{-3}$.

Janda~\etal\cite{Janda2011} studied the electron density in a self-pulsing DC transient spark discharge in atmospheric pressure air. The activity of this transient spark discharge is comparable to NRP discharges.
Using a detailed analysis of the electrical circuit they found a maximum electron density of $10^{22}$~m$^{-3}$. From Stark broadening of H$_{\alpha}$ an electron density of $10^{24}$~m$^{-3}$ is found. A reason for this discrepancy is not given.

In~\cite{Stancu2010,Stancu2010a} the densities of atomic oxygen, metastable N$_{2}$(A), N$_{2}$(B) and N$_{2}$(C) determined by LIF, cavity ring down spectroscopy (CRDS) and OES are reported in a similar NRP discharge as in~\cite{Pai2010}. They found densities in the order of $10^{24}$~m$^{-3}$ for O, $10^{21}$~m$^{-3}$ for N$_{2}$(A) and $10^{23}$~m$^{-3}$ for N$_{2}$(B) and N$_{2}$(C). However getting a complete understanding of the radical production in this discharges is difficult as the electron density, electron temperature and gas temperature are not accurately known.

The goal of this paper is to gain insight in the plasma parameters (such as the electron density and the gas temperature) of a NRP spark discharge as a first step towards understanding the plasma kinetics and radical formation. The experiments are performed in room temperature N$_{2}$ and N$_{2}$/H$_{2}$O mixtures. The effect of water is investigated since in many applications (e.g. air cleaning) water is present as impurity. Further, OH radicals can be formed due to the presence of water. OH is an interesting radical for applications such as air cleaning due to its high reactivity~\cite{Choe2006} and low selectivity. It is also of interest in combustion applications~\cite{Adamovich2009}. It additionally allows to investigate the line broadening of the hydrogen Balmer lines. In applications the plasma will be often in an air environment, however for reducing the complexity the experiments in this work are conducted in nitrogen. In future research oxygen will be added to the mixture. Time-resolved optical emission spectroscopy and Rayleigh scattering are used to obtain the electron density and gas temperature in the discharge. Significantly higher electron densities are found than in similar studies on this type of discharge~\cite{Adamovich2009,Janda2011,Laux2003,Ono2008,Pai2010,Stancu2010,Stancu2010a,Tholin2011}. The discrepancies and the different methods used will be discussed in detail.

In the next section the experimental set-up is described and the methods to determine the gas temperature and electron density are explained. In section~\ref{sec:results} the results are given. First the results from imaging are discussed, followed by the results of optical emission spectroscopy. This is followed by a discussion about the gas temperature and the electron density. Finally the plasma kinetics will be discussed before presenting the conclusions.

\section{Experimental methods}

\begin{figure}
\includegraphics[width=0.8\textwidth]{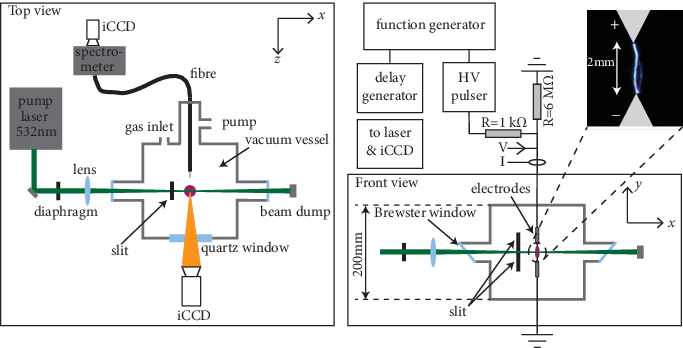} \caption{Schematic drawing of the experimental set-up and a photo of the discharge taken with a normal digital single-lens reflex camera (1 pulse).}
\label{fig:setup}
\end{figure}

The discharge is created between two needle electrodes made of tungsten. The distance between the needles is approximately 2~mm. To ensure the purity of the introduced gas the electrodes are placed in a vacuum vessel (the volume is about 8~litre), see figure~\ref{fig:setup}. The vessel is pumped down (two times) before starting the experiment to reduce impurities. During the experiment we have a continuous flow of 2.1~SLM. Water vapour is added by bubbling part of the N$_{2}$ flow through water. To calculate the water concentration it is assumed that the N$_{2}$ is saturated with water at the laboratory temperature.

High voltage pulses with a width of 170~ns, an amplitude of 9~kV and a frequency of 1~kHz are applied to the upper electrode. The pulses are created with a DEI PVX-4110 pulser triggered by a signal generator. The signal generator also triggers a delay generator which is used to time align the laser pulse, the iCCD camera and the plasma. The rise time of the HV pulse is about 80~ns (from 10-90\%), this is a bit larger than given by the manufacturer (60~ns). The drop in the voltage is significantly faster and is about 25~ns. This is because the decay time is determined by the impedance of the plasma, basically the plasma produces a short cut. The time alignment can be as good as the minimum gate width of the iCCD camera (1.2~ns according to the manual) if the plasma has no jitter. The jitter of the plasma breakdown compared to the trigger pulse is 15~ns and determines the timing accuracy of the set-up. The same set-up is used in~\cite{Verreycken2012} to perform LIF on a similar discharge in a He-H$_{2}$O mixture with a 6~ns laser pulse, accurate time alignment was achieved on this time scale due to the significantly smaller jitter for the He-plasma ignition compared to the N$_{2}$-plasma. The resistor of 6 M$\Omega$ is used to remove the charge from the charged electrode if no discharge occurs and the resistor of 1 k$\Omega$ in series with the discharge is used to limit the current through the discharge. The voltage across the electrodes and the current are measured with a voltage probe (Tektronix P6015) and current probe (Pearson 2877) as close as possible to the plasma just outside the vessel. The measured current is corrected for the capacitive current during the raising flank of the voltage pulse. To account for the difference in response time of the current (2~ns) and voltage probe (10~ns) the boxcar average of 10~ns is taken of the measured current waveform. The maximum energy delivered to the plasma is estimated by the energy which is stored in the capacity of the system. This capacity is estimated from the capacitive current and is about 25~pF, this results in a maximum input energy in the order of 1~mJ per pulse. The energy input estimated from the voltage and current waveforms is of the same order of magnitude.

To study the emission of the discharge, photos are taken with an iCCD camera (4QuikE HR camera from Stanford Computer Optics and an Andor DH534 iCCD camera, both with a Nikkor 105~mm f/4.5 UV lens ). Optical emission spectroscopy is used to determine the spectrum of the emission. The light of the discharge is collected with an optical fibre, which is placed close the plasma. The collected light is mainly coming from the bulk of the discharge, however some emission could be from the region near the electrodes. To measure time-resolved emission spectra an iCCD camera (Stanford Computer Optics 4QuikE SR) is connected to a spectrometer (27.5~cm Acton SpectraPro275). The background noise is reduced by cooling the camera with a Peltier element. Unless stated differently a gate time of 10~ns is used and a spectrum consists of 1000 accumulations. The spectra are corrected for the wavelength dependent sensitivity of the detection system by a calibrated continuum lamp (Oriel Quartz Tungsten Halogen Lamp and Ocean Optics LS-1-CAL-220).

\subsection{Gas temperature\label{sub:meth-gas}}

The rotational temperature from the N$_{2}$(C-B)(0,0) rotational band is determined by fitting experimental spectra with a program developed by Laux~\emph{et~al.}, called Specair~\cite{SPECAIR}. Due to the fast rotational energy transfer in atmospheric pressure discharges the rotational temperature is mostly equal to the gas temperature~\cite{Laux2003}. Another method which is used to determine the gas temperature is Rayleigh scattering. An EdgeWave laser (IS6II-E) produces a beam with a wavelength of 532~nm. A diaphragm and a slit are inserted to reduce the amount of scattered light originating from the electrodes, see figure~\ref{fig:setup}. A camera (4QuikE SR) is used to detect the Rayleigh signal. In front of the lens of the camera is a 532~nm interference filter which has a FWHM of about 1~nm to remove most of the plasma emission during the Rayleigh measurements. The gas density is proportional to the intensity of the Rayleigh scattering, assuming that the pressure is constant this leads to an inversely proportional dependence of the gas temperature (see also Verreycken~\etal\cite{Verreycken2011} for more details).

\subsection{Electron density\label{sub:meth-edens}}

The electron density is determined from the line broadening of spectral emission lines. In general the width of an emission line is influenced by Lorentzian (natural, resonance, Van der Waals and Stark) and Gaussian (Doppler and instrumental) broadening.

\paragraph*{Stark broadening}

For H$_{\alpha}$ the Stark broadening is calculated by Gigosos~\etal\cite{Gigosos2003}. Gigosos introduces a different reduced mass, $\mu^{*}$, which simulates a plasma which is not in thermal equilibrium ($T_{e}\neq T_{g}$): $\mu^{*}=\mu\cdot T_{e}/T_{g}$, $\mu\approx1$ is the reduced mass. For an electron temperature of 10000~K and $\mu^{*}=10$ ($T_{g}=1000$~K), the FWHM due to Stark broadening $\Delta\lambda_{Stark}$ (nm) is:
\begin{equation}
\Delta\lambda_{Stark}=8.33\cdot10^{-3}\left(\frac{n_{e}}{10^{20}}\right)^{2/3},\label{eq:Stark-H}
\end{equation}
where $n_{e}$ is the electron density (m$^{-3}$). This formula is obtained from the results in the tables of~\cite{Gigosos2003}. The broadening is 0.8~nm for $n_{e}=10^{23}$~m$^{-3}$, $T_{e}=10000$~K and $T_{g}=1000$~K. Since both the electron and gas temperature are not accurately known during the discharge, an error is induced on the electron density determined from Stark broadening. This error is estimated in section~\ref{sub:edens}.

The quadratic Stark broadening for atomic (N) lines is given by Griem~\cite{Griem1964}:
\begin{equation}
\Delta\lambda_{Stark}^{theory}=\left[1+1.75a\left(\frac{n_{e}}{10^{22}}\right)^{1/4}(1-0.75r)\right]w\left(\frac{n_{e}}{10^{22}}\right),
\end{equation}
where $a$ is the ion broadening parameter, $w$ is the electron-impact width (m) and $r$ is the Debye shielding parameter~\cite{Griem1964}. The $a$ and $w$ are tabulated in Griem~\cite{Griem1964} and are 0.035 and 4.75~pm respectively for the nitrogen line at 746~nm for an electron temperature of 10000~K. Konjevi\'{c}~\cite{Konjevic1999} has listed the ratio between the measured Stark broadening and the theoretical Stark broadening parameters listed by Griem. For the 746~nm line of N the ratio is:
\begin{equation}
\frac{\Delta\lambda_{Stark}^{meas}}{\Delta\lambda_{Stark}^{theory}}=1.197-2.97\cdot10^{-5}T_{e}.
\end{equation}
So the actual Stark broadening is:
\begin{eqnarray}
\fl\Delta\lambda_{Stark}^{meas}=\left[1.197-2.97\cdot10^{-5}T_{e}\right]\cdot\nonumber \\
\left[1+1.75a\left(\frac{n_{e}}{10^{22}}\right)^{1/4}(1-0.75r)\right]w\left(\frac{n_{e}}{10^{22}}\right).\label{eq:Stark-N}
\end{eqnarray}
For an electron density of $10^{23}$~m$^{-3}$ and an electron temperature of 10000~K, the broadening is 0.09~nm.

\paragraph*{Van der Waals broadening}

In~\cite{Djurovic2009} an expression for Van der Waals broadening $\Delta\lambda_{vd\, Waals}$ (m) is given:
\begin{equation}
\Delta\lambda_{vd\, Waals}=2.06\E{-13}\lambda_{0}^{2}\left(\alpha\bar{R^{2}}\right)^{2/5}\left(\frac{T_{g}}{\mu}\right)^{3/10}N_{n},
\end{equation}
where $\alpha$ is the polarizability of the perturber ($1.8\cdot10^{-30}$~m$^{3}$ for nitrogen gas~\cite{Bates1985}), $\mu$ is the reduced mass (kg), $N_{n}$ is the neutral number density (m$^{-3}$) and $\bar{R^{2}}$ is the difference between the values of the square radius of the emitting atom in the upper $u$ and lower level $l$. The Van der Waals broadening (nm) of H$_{\alpha}$ is:
\begin{equation}
\Delta\lambda_{vd\, Waals}=0.10\cdot\frac{p}{\left(T_{g}/300\right)^{0.7}},
\end{equation}
and for nitrogen at 746~nm:
\begin{equation}
\Delta\lambda_{vd\, Waals}=0.037\cdot\frac{p}{\left(T_{g}/300\right)^{0.7}},
\end{equation}
where $p$ is the pressure (bar). At room temperature and atmospheric pressure this results in a broadening of 0.1~nm for H$_{\alpha}$ and 0.04~nm for N.

The instrumental broadening Gaussian FWHM is measured with a low pressure mercury/argon Pen-Ray light source (line width $\approx1$~pm) and is 0.19~nm.

Natural, resonance and Doppler broadening ($T_{g}<1000$~K, see further) are smaller than 5~pm and can therefore be neglected in our conditions~\cite{Laux2003}.

\paragraph*{Fitting procedure}
Since the width of the H$_{\alpha}$ line is always larger than 1~nm and the line is well fitted by a Lorentzian profile, the electron density is obtained by $\Delta\lambda_{Stark}=\Delta\lambda_{meas}-\Delta\lambda_{vd\, Waals}$.

The broadening of N is smaller and the instrumental broadening becomes significant in the recombination phase. In this case the deconvolution to obtain the Stark FWHM is made by the following approximative formula~\cite{Bruggeman2009a}:
\begin{eqnarray}
\Delta\lambda_{meas} & = & \frac{\Delta\lambda_{Stark}+\Delta\lambda_{Vd\, Waals}}{2}+\nonumber \\
 &  & \sqrt{\frac{\left(\Delta\lambda_{Stark}+\Delta\lambda_{Vd\, Waals}\right)^{2}}{4}+\Delta\lambda_{instr}^{2}}.\label{eq:vdW-and-Stark}
\end{eqnarray}

\section{Results and Discussion\label{sec:results}}

\subsection{Emission}

\begin{figure}
\hfill{}\subfigure{%
\begin{minipage}[c]{0.6\textwidth}%
\includegraphics[width=1\textwidth]{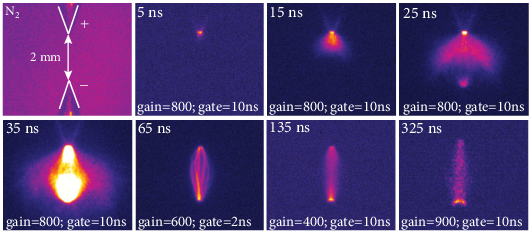}%
\end{minipage}}
\hfill{}
\subfigure{%
\begin{minipage}[c]{0.35\textwidth}%
\includegraphics[width=1\textwidth]{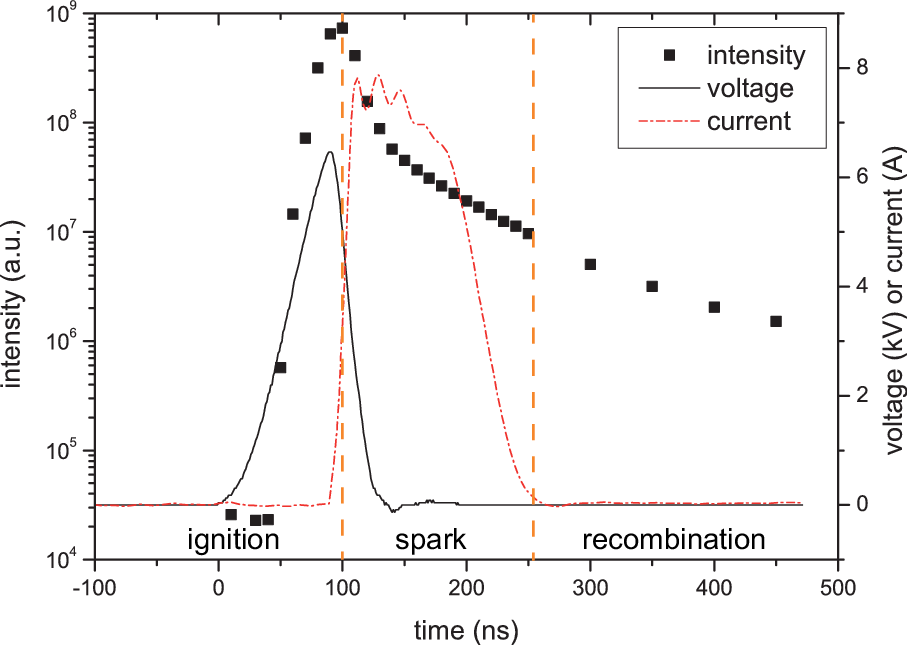}%
\end{minipage}}
\hfill{}
\caption{(a) Images of the emission of the nitrogen plasma. The emission is accumulated over 1000 discharges, except for $t=65$~ns only 2 discharges are accumulated. Note that the gain is not kept constant so no conclusion on relative intensities can be made. The image at $t=35$~ns is overexposed, however it clearly shows the evolution of the discharge. The pictures are made with the 4QuikE camera. (b) Relative intensity of the emission, which is rescaled to a gain of 900.}
\label{fig:images}
\end{figure}

The discharge can be divided into three phases. The voltage across the electrodes increases during the first 100~ns, while the current is small (see figure~\ref{fig:images}b). This is the \emph{ignition phase} of the discharge. In this phase an ionization channel is formed between the electrodes. If the conductivity of this channel is high enough, the voltage collapses and the current increases up to a few Amps and a spark is formed (\emph{spark phase}). After 200~ns the current decreases as the power source is switched off and plasma recombination starts (\emph{recombination phase}).

Time-resolved images of the discharge are shown in figure~\ref{fig:images}a. The images are an accumulation of 1000 discharges, except the figure at $t=65$~ns, this is an accumulation of 2 discharges to study the filamentation. Since the gate time (10~ns) is smaller than the plasma jitter (15~ns), an overlap between the images in figure~\ref{fig:images}a is not excluded. However, if we look at the images at 25~ns and 35~ns, the influence seems marginal since we do not see any emission of the (saturated) image at 35 ns in the image of 25 ns. Furthermore the mentioned jitter is on the breakdown of the discharge, so before 100~ns the resolution is better than 15 ns. In other studies (e.g.\ \cite{Nistor2008}) the gain of the iCCD camera is kept constant. However, to be able to study the morphology of the discharge, the gain cannot be kept constant in our case since the emission intensity changes more than 4 orders of magnitude (see figure~\ref{fig:images}b).

From figure~\ref{fig:images}a one can see that the discharge starts at the (upper) positive electrode. At the anode a cloud is formed which travels towards the cathode. This behaviour has also been observed previously by e.g.\ Briels~\cite{Briels2008} at reduced pressure. The emission of the discharge is still low at this point (see figure~\ref{fig:images}b).

Experiments and simulations on the emission of a NRP streamer/glow discharge in a pin-pin electrode configuration are performed in~\cite{Tholin2011,Celestin2009,Sobota2010}. Here it is also observed that the discharge starts at the anode and moves towards the cathode. Further they found that at the start of the discharge also some emission is present at the cathode, however it is less intense than at the anode. From the model, this emission is explained by the development of a weak negative streamer~\cite{Celestin2009}. In our case this emission cannot be observed directly at the start of the discharge, however at 25~ns (close before the positive and negative streamers connect) we indeed observe emission at the cathode side.

After 35~ns filaments start to develop, at $t=65$~ns multiple filaments are visible in figure~\ref{fig:images}a. The filaments indicate that the discharge is constricted, which results in an increase in the current. This increase in the current cannot be seen in figure~\ref{fig:images}b, probably because the increase is too small to be measured by the 8-bit digitizer of the oscilloscope. At this point the intensity of the discharge has increased by several orders of magnitude (figure~\ref{fig:images}b).

\begin{figure}
\includegraphics[width=0.8\textwidth]{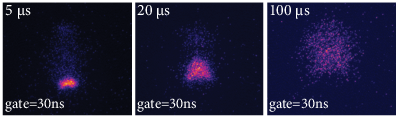}
\caption{Images of the emission of the nitrogen plasma at microsecond time scales. The emission is accumulated over 1000 discharges. The images are made with the Andor camera at maximum gain. \label{fig:images_us}}
\end{figure}

The emission is homogeneously distributed during and after the current pulse (after 100~ns). To check whether this homogeneous distribution is caused by the accumulation of 1000 discharges, images were taken after 100~ns where only 2 discharges were accumulated (not shown in figure~\ref{fig:images}a). These pictures also show a homogeneous distribution. The most intense emission is coming from the region near the cathode and seems to shift towards the bulk during the afterglow on a time scale of tens of microseconds (see figure~\ref{fig:images_us}). This indicates that a gradient in the density of metastable or the ionic species (electrons) is present.

The images of the discharge in N$_{2}$/0.9\% H$_{2}$O are very similar and therefore not presented.

\begin{figure}
\includegraphics[width=0.95\textwidth]{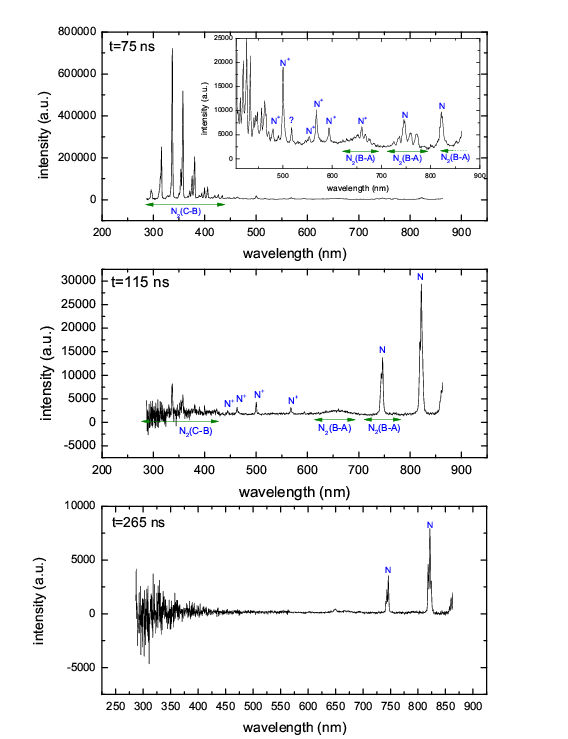}
\caption{Overview spectra of a N$_{2}$ plasma during the three phases. Each spectrum consists of 1000 accumulations. The upper image is the spectrum during the ignition phase of the discharge, the middle during the spark phase and the bottom spectrum of the recombination phase. The lines are identified with the use of Lofthus~\cite{Lofthus1977} and the NIST database~\cite{NIST_lines}.}
\label{fig:spectra}
\end{figure}

The overview spectra of pure nitrogen are given in figure~\ref{fig:spectra}. Due to the relative low sensitivity of the spectrometer at shorter wavelengths, the noise between 300 and 400~nm is relatively large in the spark and recombination phase.

The spectra show clearly the emission from N$_{2}$(C-B), second positive system (SPS), in the range 300-450~nm and the N$_{2}$(B-A), first positive system (FPS), in the range of 600-850~nm. The N$_{2}^{+}$(B-X), first negative system (FNS), has a spectral band at 391~nm, which is clearly visible in the spectrum recorded with a higher resolution. Other lines that are observed originate from excited N and N$^{+}$. Emission from O impurities (the emission line at 777~nm) is not observed.

The emission is very similar when water is added, this is because the concentration of water ($10^{23}$~m$^{-3}$) is smaller or comparable to the electron density (see further). Therefore the plasma emission is still dominated by nitrogen. There are also some differences if water is added, the main differences are discussed below. First of all the emission of molecular nitrogen, compared to atomic emission, is less strong when water is added to the nitrogen gas. The addition of 0.9\% of water increases the theoretical quenching rate of N$_{2}$(C) by 30\% (so the lifetime of N$_{2}$(C) goes from 3.4~ns to 27~ns)~\cite{Pancheshnyi2000}. Due to the quenching effect of water the excited nitrogen states are more lost without radiation. This causes the decreased intensity of molecular nitrogen emission. The first positive system of nitrogen is not visible at all in this case.

Due to the addition of water, emission of hydrogen and NH(A-X) become visible. The H$_{\beta}$ line is only visible during the ignition phase of the discharge. The wings of the H$_{\alpha}$ line are overlapped with the N$^{+}$ lines, which is taken into account in the fitting procedure. During the ignition phase the N$^{+}$ lines are too strong compared to the H$_{\alpha}$ line to discern the H$_{\alpha}$ line.

\subsection{Gas temperature\label{sub:Tgas}}

The gas temperature can be determined from the SPS of nitrogen only in the ignition phase, since only in this phase molecular nitrogen emission is visible. The measured gas temperature between 65~ns and 95~ns after the start of the voltage pulse is constant and is $(350\pm50)\,\mbox{K}$ in pure N$_{2}$. For the Rayleigh measurements no significant dip in the intensity is observed in the ignition phase, which indicates that the gas temperature is indeed close to room temperature.

\begin{figure}
\includegraphics[width=0.8\textwidth]{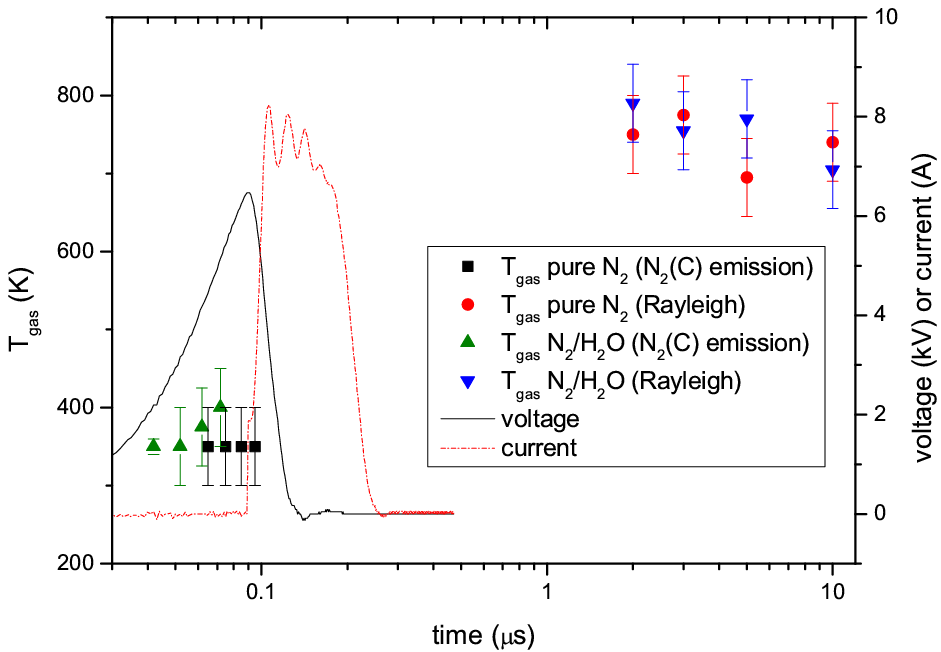}
\caption{The gas temperature of the discharge as function of time in pure N$_{2}$ and N$_{2}$/0.9\% H$_{2}$O. The voltage and current waveforms are given as reference.}
\label{fig:Tgas}
\end{figure}

After the discharge is switched off the temperature is determined by Rayleigh scattering, see figure~\ref{fig:Tgas}. During the spark phase of the discharge the gas temperature could not be accurately measured since the emission intensity of the discharge is too high compared to Rayleigh signal in spite the fact that a bandpass filter was used. During the recombination phase of the discharge the temperature is about 750~K. This could be due to heating of the gas when the plasma is on by elastic collisions between electrons and N$_{2}$. If all the energy of the applied voltage pulse (1~mJ) is converted into gas heating of the plasma volume, the gas temperature would be 1100~K. However, the impulse transfer rate of elastic collisions between electrons and neutrals is about $k_{elas}=6\E{-14}$~m$^{3}$\,s$^{-1}$~\cite{BOLSIG}. The rate of energy transfer is $k_{energy}=2m_{e}k_{elas}/m_{neutrals}\approx2\cdot10^{-18}$~m$^{3}$\,s$^{-1}$. With an electron density of the order of $10^{24}$~m$^{-3}$ (see also further), the energy transfer time is about 0.5~\textmu{}s. Note that at such high electron densities Coulomb collisions are dominant and elastic collisions will contribute marginally to the plasma heating.

The elevated temperature at large time scales (10~\textmu{}s) has also been observed by Ono~and~Oda~\cite{Ono2008} in pulsed corona discharges in a H$_{2}$O/O$_{2}$/N$_{2}$ mixture. They explained the elevated temperature by energy transfer from vibrationally excited molecules to kinetic gas energy (vibrational to translational energy transfer V-T), which is fast if O$_{2}$ is present. The V-T rate for nitrogen and water is rather small and too slow to explain the heating in the current study~\cite{Fridman2008}.

Mintoussov\emph{~et~al.}~\cite{Mintoussov2011} also observed an elevated temperature after the discharge. Their experiments have been performed in air at low pressure (\textless 10~mbar). They explained the heating of the gas by quenching of excited molecular nitrogen and electron-ion recombination processes of N$_{2}^{+}$. The time scales for these processes are in the order of 10~ns. Via several collisions the energy is then transferred to kinetic energy. This can explain the elevated temperature in the current experiment, especially in view of the large electron density.

The gas temperature in a N$_{2}$/H$_{2}$O discharge is similar within the margin of error.

\subsection{Electron density\label{sub:edens}}

\begin{figure}
\hfill{} \subfigure[Line broadening of the nitrogen line at 746~nm in a pure nitrogen discharge at 125~ns after the start of the voltage pulse.]{\includegraphics[width=0.455\textwidth]{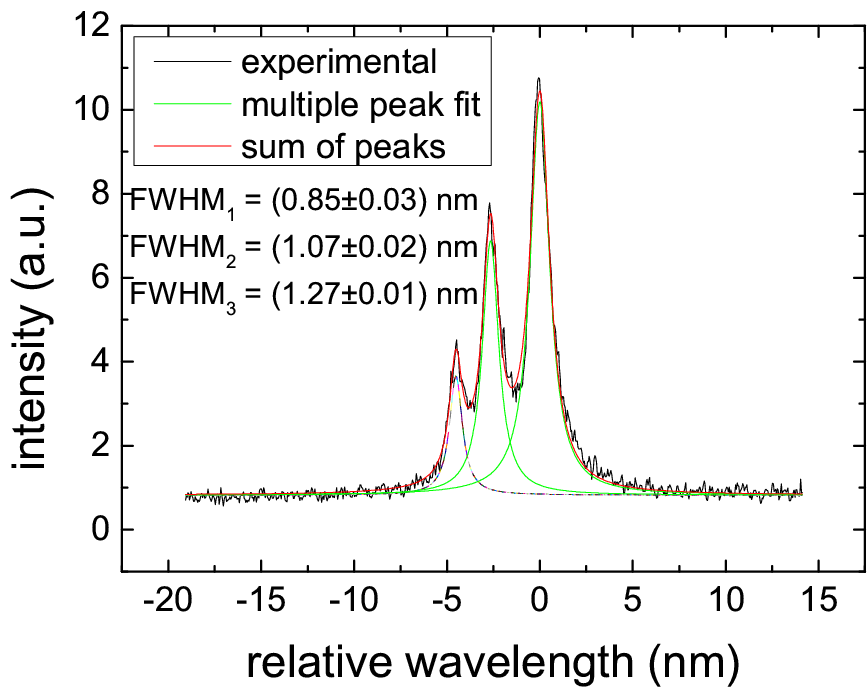}}
\hfill{} \subfigure[Line broadening of the H$_{\alpha}$ in a nitrogen with 0.9\% water vapour discharge at 122~ns after the start of the voltage pulse. The spectrum of N is subtracted from the spectrum of H$_{\alpha}$ and N, to obtain only the H$_{\alpha}$ line.]{\includegraphics[width=0.45\textwidth]{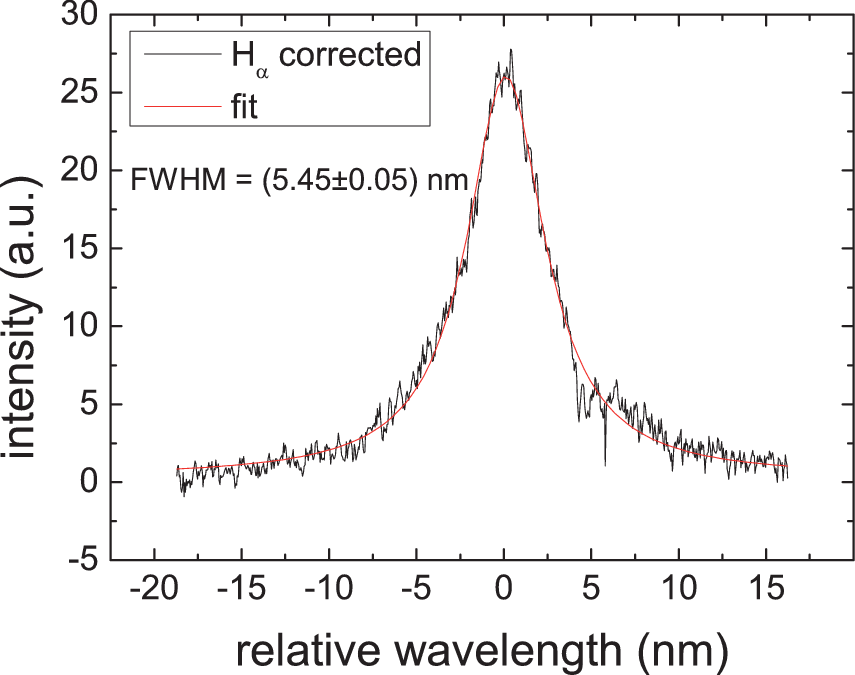}}
\hfill{}
\caption{Examples of fits (Lorentzian profile) of the N- and H$_{\alpha}$-line spectra.}
\label{fig:linebroadening}
\end{figure}

\begin{figure}
\includegraphics[width=0.8\textwidth]{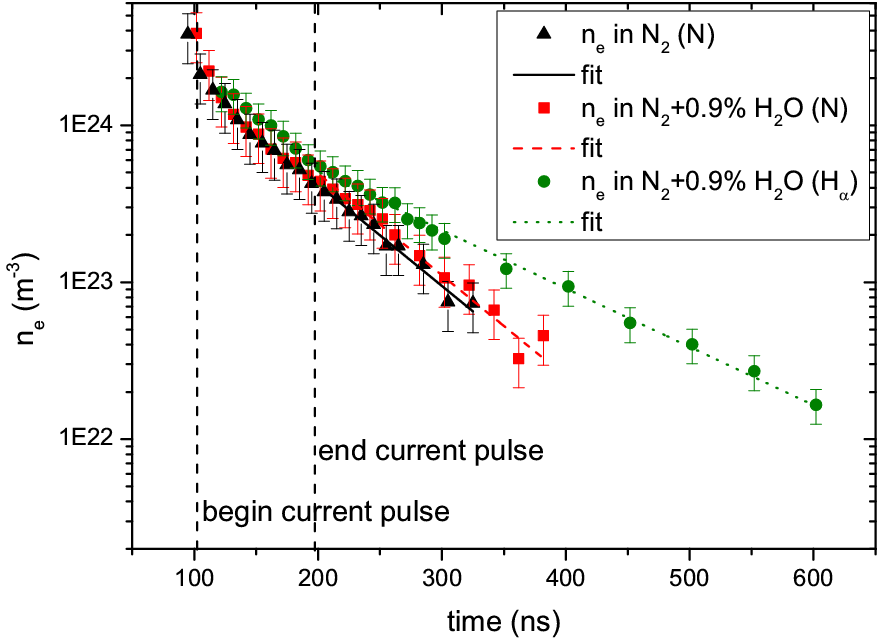}
\caption{The electron density versus time in a N$_{2}$ discharge and a N$_{2}$ with 0.9\% H$_{2}$O discharge. }
\label{fig:edens}
\end{figure}

The electron density is estimated from the line width of a N line at 746~nm and the H$_{\alpha}$-line as described in section~\ref{sub:meth-edens}. In figure~\ref{fig:linebroadening} are two examples of the line broadening of the N and H$_{\alpha}$ lines at about 125\,ns. The measured line is fitted with a Lorentzian line shape. The FWHM are indicated in the figures. The electron density as function of time is given in figure~\ref{fig:edens}. A typical value of the electron temperature in a thermal atmospheric pressure plasma with a high electron density is 15000~K~\cite{Boulos1994}. Due to reactions of electrons with vibrationally excited N$_{2}$ in the afterglow the electron temperature follows normally the vibrational temperature rather than dropping immediately to the gas temperature~\cite{Guerra2004,Gorbunov2001,Ambrico2005}. Therefore an electron temperature of 10000~K is assumed in the spark and recombination phase and since $T_{e}>T_{g}$, 2500~K is taken as a lower estimate of $T_{e}$. The gas temperature is about 1000\,K, see section~\ref{sub:Tgas}.

To estimate the error in the electron density determined from H$_{\alpha}$ due to the assumption that $T_{e}=10000$~K, the electron density has also been determined for $T_{e}=2500$~K ($T_{g}=1000\,$K) and $T_{e}=20000$~K ($T_{g}=2000\,$K, as no information in~\cite{Gigosos2003} about $T_{e}=20000$~K and $T_{g}=1000\,$K is available). Also the gas temperature is not known in this time interval where the electron density is determined, to estimate this error the electron density is determined for $T_{g}=2000$~K ($T_{e}=10000$~K). Both effects result in a total error of not more than 25\%. The same procedure has been used to estimate the error on the electron density determined for the N-line. The error due to $T_{e}$ is in this case 35\%. There is no data available to estimate the error induced by the uncertainty in $T_{g}.$

The measured electron densities in a pure N$_{2}$ plasma and a N$_{2}$/0.9\% H$_{2}$O discharge are of the order $10^{24}$~m$^{-3}$ and correspond within the margin of error. The highest electron density is found, within the margin of error due to the jitter, at the start of the current pulse. The maximum electron density is $4\cdot10^{24}$~m$^{-3}$. The maximum ionization degree is 50\%, assuming a gas temperature of 1000~K and a pressure of 1 bar in the spark phase. Due to the discharge the pressure in the plasma can increase to about 3 bar (no gas expansion), this results in a maximum ionization rate of 16\%. So the ionization degree local in the filament during a few tens of nanoseconds is 16-50\%, which is unusually high.

To check the assumption that the broadening is mainly due to the Stark effect, the contribution of additional broadening mechanisms is estimated. First the influence of an external electrical field is estimated. The broadening of H$_{\alpha}$ due to an external electrical field (assuming that the maximum E-field is 80~kV\,cm$^{-1}$) is negligible compared to the observed line profile~\cite{Gonzalez}, it is assumed that this also holds for the atomic nitrogen line.

Another possible additional contribution to the broadening is that the Van der Waals broadening is underestimated since it is assumed that the pressure remains constant at 1~bar. Due to the discharge the pressure in the plasma channel can increase for a short time, thereby increasing the Van der Waals broadening. However to explain the observed broadening only by Van der Waals broadening a pressure of 60~bars is needed. When the temperature increases from 350~K to 1500~K and there is no expansion of the plasma channel, the pressure increases by a factor of 5 at maximum. So this effect contributes less than 5\% to the total broadening.

A possible broadening mechanism which has not been taken into account is self-absorption. As the discharge has a diameter of only 0.2~mm, the effect of self-absorption on the electron density is less than 5\% for N (assuming the N density equals $2.5\E{25}$~m$^{-3}$). For H$_{\alpha}$ the effect of self-absorption is even smaller.

From the electron density $n_{e}$ (m$^{-3}$) and current density $J$ (A\,m$^{-2}$), the electrical field during the spark phase can be estimated. The current density is:
\begin{equation}
J=\frac{I}{A}=\sigma E=\frac{n_{e}e^{2}}{m_{e}\nu_{c}}E,\label{eq:current}
\end{equation}
where $I$ is the current (A), $A$ is the area through which the current flows (m$^{2}$), $\sigma$ is the conductivity ($\Omega^{-1}$\,m$^{-1}$), $\nu_{c}$ is the collision frequency (s$^{-1}$) and $E$ is the electrical field (V\,m$^{-1}$). In a plasma with a high ionization rate the collisions are dominated by electron-ion collisions and the collision frequency is then~\cite{Bittencourt2004}:
\begin{equation}
\nu_{c}=\nu_{ei}=\frac{n_{e}e^{4}\ln\Lambda}{18\sqrt{2}\pi\varepsilon_{0}^{2}\sqrt{m_{e}}(k_{B}T_{e})^{3/2}},
\end{equation}
where $\ln\Lambda$ is the Coulomb integral~\cite{Bittencourt2004}:
\begin{equation}
\ln\Lambda=\ln\left(\frac{12\pi(\varepsilon_{0}k_{B}T_{e})^{3/2}}{n_{e}^{1/2}e^{3}}\right).
\end{equation}
The current at $t=140$~ns is approximately 8~A, the plasma width is 0.2~mm and the electron density about $10^{24}$~m$^{-3}$. The electrical field obtained from the above formula is about $30$~kV\,m$^{-1}$, corresponding to a voltage across the electrodes of approximately 65~V. This is consistent with the measurement, see figure~\ref{fig:TR-OES}, which indicates that the voltage is smaller than 100~V. To check the sensitivity of the electron density to the voltage, we assume that the electron density is 100 times lower. The voltage across the electrodes is then approximately 100~V, which is still consistent with the measurement (see figure~\ref{fig:TR-OES}). If the electron density is high it is only present in equation~\ref{eq:current} in the logarithm ($\ln\left(n_{e}^{-1/2}\right)$), so the current density is rather insensitive to changes in the electron density. Hence we can conclude that, at high electron densities, current density measurements are not accurate enough to determine the electron density.

From all of the above, we can conclude that the measured broadening is dominated by Stark broadening.

The obtained electron densities are significantly larger than found in literature by conductivity (current) measurements of very similar discharges~\cite{Pai2010,Janda2011}. Pai~\etal\cite{Pai2010} determined the electron density from the plasma current. In their calculations only electron-atom collisions are taken into account. The neglect of electron-ion collisions could result in an underestimation of the electron density. However it must be noted that the plasma on-time is smaller and as a consequence the electron density does not reach the large values as in the current study. The line broadening of H$_{\alpha}$ reported by Janda~\etal\cite{Janda2011} is very similar to the one observed in the discharge presented in the current paper. Janda~\etal\cite{Janda2011} also determined the electron density from the conductivity $\sigma_{p}$ of the plasma:
\begin{equation}
n_{e}\propto\sigma_{p}\cdot\nu_{c},
\end{equation}
where $\nu_{c}$ is the collision frequency. Janda~\emph{et~al.} however used the collision frequency for discharges with a small ionization degree, i.e.~\cite{Janda2007}. However if the ionization degree is larger than $10^{-3}$, the Coulomb collisions between electrons and ions start to dominate. This is a reason for underestimation of the collision frequency, and thus the electron density.

Note that when estimating the energy necessary to obtain this large electron density with the plasma volume calculated from the FWHM of the plasma (200~\textmu{}m), you obtain an energy in the same order of magnitude as the estimate of input energy. However the filamentation of the discharge (see figure~\ref{fig:images}) causes that the effective value in which the discharge reaches this large electron density is significantly smaller than computed from this estimate of the volume. This allows indeed for the additional losses in energy by (vibrational) excitation and dissociation.

\subsection{Plasma kinetics}

\begin{figure}
\includegraphics[width=0.8\textwidth]{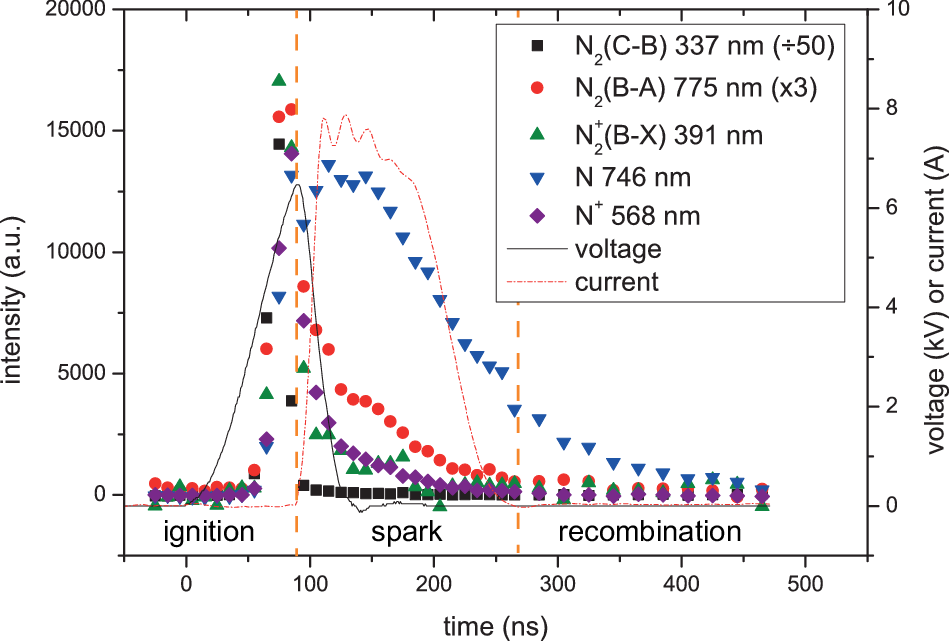}
\caption{Maximum intensity of several spectral lines and bands as function of time in a N$_{2}$ plasma. The peak intensity of the N$_{2}$(C-B) line at 337~nm is divided by 50 and the intensity of the N$_{2}$(B-A) at 775~nm is multiplied by 3. To account for the difference in response time of the current (2~ns) and voltage probe (10~ns) the boxcar average of 10~ns is taken of the measured current waveform.
\label{fig:TR-OES} }
\end{figure}

To study the plasma kinetics the peak intensity of the most intense spectral lines as function of time is shown in figure~\ref{fig:TR-OES}. In the ignition phase of the discharge the emission is coming from the FPS, SPS and FNS of nitrogen. Due to the high electron temperature, the excited states N$_{2}$(C), N$_{2}$(B) and N$_{2}^{+}$(B) are created by direct electron impact. Note that N$_{2}$(C) can also be created via a pooling reaction if the metastable density is large~\cite{Akishev2010}:
\begin{equation}
\mbox{N}_{2}(\mbox{A})+\mbox{N}_{2}(\mbox{A})\too\mbox{N}_{2}(\mbox{X})+\mbox{N}_{2}(\mbox{C}),
\end{equation}
and N$_{2}^{+}$(B) can also be created by vibrational pumping~\cite{Linss2004}:
\begin{equation}
\mbox{N}_{2}^{+}(\mbox{X})+\mbox{N}_{2}(\mbox{X},\,\nu\geq12)\too\mbox{N}_{2}^{+}(\mbox{B})+\mbox{N}_{2}(\mbox{X},\,\nu<12),
\end{equation}
however, the time scale for these reactions are \textgreater{}0.5~\textmu{}s if the ion and metastable densities are smaller than $10^{22}$~m$^{-3}$~\cite{Akishev2010,Bruhl1997}. Short (30~ns) after the start of the emission of the FPS the emission of atomic and ionic nitrogen lines starts and the emission of molecular nitrogen collapses. The excitation energy of N is larger than of N$_{2}$, so a drop in the electron temperature cannot explain this behaviour. Clearly the concentration of the atomic nitrogen needs to increase due to dissociation processes, since dissociative recombination of N$_{2}^{+}$ yields only 5.8~eV excess energy and can therefore not produce excited nitrogen atoms (see table~\ref{tab:Nproduction}). This is consistent with the large electron density at this time (figure~\ref{fig:edens}).

During the spark phase of the discharge there is no significant molecular emission. In figure~\ref{fig:TR-OES} it seems that there is emission of the FPS of nitrogen in the spark phase, however this could be partly caused by an overlap between the band of the FPS and atomic nitrogen lines. In the recombination phase only emission of atomic nitrogen lines is visible. This clearly shows that these excited states are produced by recombination processes.

From figure~\ref{fig:TR-OES} the decay time of N in the recombination phase can be determined by fitting the curve with an exponential function starting at the time when the discharge is switched off. The experimental decay times of N emission at 746~nm and 822~nm are respectively ($92\pm3)$~ns and ($82\pm3$~ns). According to the NIST database~\cite{NIST_lines} the lifetimes are respectively 51~ns and 44~ns, these lifetimes will be even shorter due to quenching. So, even considering the 15~ns jitter, there is a significant production of excited atomic nitrogen after the discharge is switched off. There are several processes which can create atomic nitrogen in the recombination phase when $T_{e}$ is low. The reactions are listed in table~\ref{tab:Nproduction}. The reactions should have enough energy to create an excited N atom in the upper level of the 746~nm and 822~nm transition, which is 12~eV and 11.8~eV respectively~\cite{NIST_lines} . Only reactions R1 till R3 have enough excess energy.

\begin{table}
\caption{Processes which can create atomic nitrogen in the ground state. The reaction rates and the excess energy of the reactions are also shown~\cite{Akishev2010}.}
\begin{tabular}{@{}llll}
\br
No  & Reaction  & Rate constant (m$^{6}$\,s$^{-1}$, m$^{3}$\,s$^{-1}$ )  & Excess energy (eV)\\
\mr
R1  & $\mbox{N}^{+}+e\too\mbox{N}+h\nu$  & $3.4\E{-16}T_{e}^{-3/4}$ \footnotemark[1]  & $<14.5$\\
R2  & $\mbox{N}^{+}+2e\too\mbox{N}+e$  & $7\E{-32}(300/T_{e})^{4.5}$  & $14.5$\\
R3  & $\mbox{N}^{+}+e+\mbox{N}_{2}\too\mbox{N}+\mbox{N}_{2}$  & $6\E{-39}(300/T_{e})^{1.5}$  & 14.5\\
R4  & $\mbox{N}_{2}^{+}+e\too2\mbox{N}$  & $1.8\E{-13}(300/T_{e})^{0.4}$  & 5.8\\
R5  & $\mbox{N}_{3}^{+}+e\too\mbox{N}_{2}+\mbox{N}$  & $1.2\E{-12}(300/T_{e})^{0.5}$  & 11.1\\
\br
\end{tabular}
\label{tab:Nproduction}
\begin{itemize}
\item[$\dagger$] From~\cite{Fridman2008}.
\end{itemize}
\end{table}

To determine which reactions are dominant, the reaction rates have to be compared. Assuming that the electron temperature in the afterglow is $T_{e}=10000\,\mbox{K}$ (see section~\ref{sub:edens}), the rate constants for reactions R1, R2 and R3 are $3.4\E{-19}$~m$^{3}$\,s$^{-1}$, $9.8\cdot10^{-39}$~m$^{6}$\,s$^{-1}$ and $3.1\cdot10^{-41}$~m$^{6}$\,s$^{-1}$ respectively~\cite{Akishev2010}. To compare the rates of R2 and R3 with the rate of equation R1, the rates are multiplied by the electron density and nitrogen density respectively. The electron density is in the order of $10^{24}$~m$^{-3}$, see section~\ref{sub:edens}. The nitrogen density at 750~K and 1~atm is $9.8\cdot10^{24}$~m$^{-3}$. This results in rate coefficients of $9.8\cdot10^{-15}$~m$^{3}$\,s$^{-1}$ and $3.0\cdot10^{-16}$~m$^{3}$\,s$^{-1}$ for R2 and R3 respectively. Therefore reaction R2 is dominant. Note that the rates for charge exchange:
\begin{eqnarray}
 &  & \mbox{N}^{+}+2\mbox{N}_{2}\too\mbox{N}_{3}^{+}+\mbox{N}_{2}\\
 &  & \mbox{N}^{+}+\mbox{N}_{2}\too\mbox{N}_{2}^{+}+\mbox{N}
\end{eqnarray}
(respectively $2.5\E{-16}$~m$^{3}$\,s$^{-1}$ and $10^{-18}$~m$^{3}$\,s$^{-1}$~\cite{Akishev2010}) and subsequent dissociative recombination (R4 and R5 in table~\ref{tab:Nproduction}) are too slow to compete with reaction R2 to cause the $n_{e}$ decay.

These results indicate that atomic nitrogen is still produced in the recombination phase due to 3 body electron-ion recombination reactions in the recombination phase.

The measured decay time of the N emission (line at 746~nm) if water is added to the gas mixture is $(120\pm3)$~ns. So the decay time is about 30\% larger than in pure nitrogen. Therefore the source term should be faster if water is added. The reaction R2 in table~\ref{tab:Nproduction} is the dominant source term for excited N atoms, as mentioned before. The gas temperature does not change when water is added, see section~\ref{sub:Tgas}. A reason could be that the electron density is slightly larger if water is added, see section~\ref{sub:edens}. This will increase the production rate of N via reaction R2. The difference in the electron density is however very small (see figure~\ref{fig:edens}).

The decay of the electron density can be determined from figure~\ref{fig:edens}. It can bee seen that the decay is linear on the logarithmic scale, so the electron density is governed by the following differential equation:
\begin{equation}
\frac{dn_{e}(t)}{dt}=-kn_{e}(t)\Longrightarrow n_{e}=n_{e0}\exp\left[-k_{1}\left(t-t_{0}\right)\right].
\end{equation}
where $k_{1}$ is the reaction rate (s$^{-1}$). This results in a rate constant $k_{1}=(1.49\pm0.08)\E{7}$~s$^{-1}$ in a pure N$_{2}$ discharge. The time scale for diffusion $\tau=(R/2.4)^{2}/D_{a}$, where $R=0.1$~mm is the radius of the plasma channel and $D_{a}\approx(1+T_{e}/T_{i})D_{i}$ the ambipolar diffusion coefficient. The electron temperature is at maximum 20 times the ion temperature $T_{i}$. The ion diffusion coefficient $D_{i}$ can be calculated from the ion mobility $\mu_{i}$ which can be found in~\cite{Ellis1976} for the N$^{+}$-ion in N$_{2}$, $D_{i}=k_{B}T_{i}\mu_{i}/e=2.5\E{-5}$~m$^{2}$\,s$^{-1}$. So the fastest possible time scale for ambipolar diffusion is 3.5~\textmu s, which is 2 orders of magnitude slower than the decay time of the electron density.

Based on the reaction rates (see table~\ref{tab:Nproduction}), one would expect that the electron density decays due to electron-ion recombination, which gives the following decay curve (assuming that the ion density is equal to the electron density):
\begin{equation}
\frac{dn_{e}(t)}{dt}=-k_{2}n_{e}^{3}(t)\Longrightarrow n_{e}(t)=\frac{n_{e0}}{\sqrt{1+2k_{2}n_{e0}^{2}\left(t-t_{0}\right)}},
\end{equation}
where $k_{2}$ is the reaction rate for electron-ion recombination (m$^{6}$\,s$^{-1}$). The experimental decay could however not be fitted with this decay curve. According to literature $k_{2}=9.8\E{-39}$~m$^{6}$\,s$^{-1}$ at $T_{e}=10000$~K~\cite{Akishev2010}. At the end of the voltage pulse the decay rate would then be $2\E{9}$~s$^{-1}$ ($n_{e}=5\E{23}$~m$^{-3}$), which is more than 2 orders of magnitude faster than the measured rate. As mentioned above, the jitter is about 15~ns. This results in an error in the decay time of maximum a factor 10, so this cannot explain the slow decay time of the electron density. Another possible explanation is that the electron temperature is higher than 10000~K. In order to explain the difference the electron temperature should be 2.3~eV, this is however large for a recombining plasma. This indicates that there is a significant production of electrons during the recombination phase. One hypothesis is that electrons are created by Penning and associative ionization reactions~\cite{Linss2004,Starikovskaia2001,Akishev2010}, see table~\ref{tab:Penning-and-associative}. These ionization processes should take place on a time scale in the order of 100~ns. This means that the density of N$_{2}$(a$^{\prime}$) should be, considering the different rates as found in literature (see table~\ref{tab:Penning-and-associative}), in the order of $5\E{22}-10^{24}$~m$^{-3}$ and the density of N($^{2}$P$^{0}$) should be larger than $2\E{24}$~m$^{-3}$, which is in the same order of magnitude as the electron density, so rather realistic. N$_{2}$(a$^{\prime}$) can for instance be created by recombination of two N($^{2}$D$^{0}$) metastables~\cite{Lofthus1977}. It clearly indicates the large amount of energy which is stored in dissociated N and nitrogen metastables.

\begin{table}
\caption{Penning and associative ionization reactions and their reaction rates.}
\begin{tabular}{llll}
\br
No.  & Reaction  & Reaction rate (m$^{-3}$\,s$^{-1}$)  & Reference\\
\mr
R6  & $\mbox{N}_{2}(\mbox{a}^{\prime})+\mbox{N}_{2}(\mbox{a}^{\prime})\too\mbox{N}_{2}\mbox{(X)}+\mbox{N}_{2}^{+}\mbox{(X)}+e$  & $5\E{-17}$  & \cite{Starikovskaia2001}\\
R7  & $\mbox{N}_{2}(\mbox{a}^{\prime})+\mbox{N}_{2}(\mbox{a}^{\prime})\too\mbox{N}_{4}^{+}+e$  & $10^{-17}$  & \cite{Akishev2010}\\
 &  & $5\E{-17}$  & \cite{Akishev2007}\\
 &  & $2\E{-16}$  & \cite{Starikovskaia2001,Akishev2010}\\
R8  & $\mbox{N}(^{2}\mbox{D}^{0})+\mbox{N}(^{2}\mbox{P}^{0})\too\mbox{N}_{2}^{+}\mbox{(X)}+e$  & $10^{-19}$  & \cite{Akishev2007}\\
 &  & $2\E{-18}$  & \cite{Starikovskaia2001}\\
R9  & $\mbox{N}(^{2}\mbox{P}^{0})+\mbox{N}(^{2}\mbox{P}^{0})\too\mbox{N}_{2}^{+}\mbox{(X)}+e$  & $5\E{-18}$  & \cite{Starikovskaia2001}\\
\br
\end{tabular}
\label{tab:Penning-and-associative}
\end{table}

The decay rate of $n_{e}$ determined from the N line is $(1.49\pm0.08)\E{7}$~s$^{-1}$. The decay rate corresponds within the margin of error to the decay rate measured in a pure nitrogen discharge. The decay rate determined from the H$_{\alpha}$ line is $(8.6\pm0.2)\E{6}$~s$^{-1}$, which is very similar to the decay rate determined from the N line.

\section{Conclusion}

Nanosecond pulsed discharges in N$_{2}$ and N$_{2}$/0.9\% H$_{2}$O are studied with time-resolved optical emission spectroscopy and Rayleigh scattering. The evolution of the discharge consists of three phases, the ignition phase (raising voltage, no current), the spark phase (low voltage, high current) and the recombination phase. From time-resolved imaging it is observed that the discharge starts at the positive electrode during the ignition phase. At the end of the ignition phase multiple filaments are visible. When the spark phase starts, a homogeneous plasma channel is visible.

During the ignition phase the emission is mainly coming from molecular nitrogen. This emission of molecular nitrogen collapses at the end of this phase and the emission of atomic and ionic nitrogen increases. At the start of the spark phase only atomic nitrogen emission is visible. Due to the quenching of molecular nitrogen by water the emission of molecular nitrogen is weaker in the discharge with added water vapour. The atomic emission is largely caused by recombination reactions.

From the spectrum of molecular nitrogen the temperature during the ignition phase is estimated to be 350~K. In the recombination phase (1~\textmu{}s after the start of the discharge) the temperature has been estimated with Rayleigh scattering and is about 750~K. The addition of 0.9\% water has no significant influence on the gas temperature.

The electron density has been determined with the broadening of an atomic nitrogen line and H$_{\alpha}$ in the N$_{2}$/H$_{2}$O mixture. Electron densities up to $4\E{24}$~m$^{-3}$ have been measured in a pure N$_{2}$ and N$_{2}$/H$_{2}$O discharge. During the recombination phase the electron density decreases with a rate constant in the order of $10^{7}$~s$^{-1}$ for both $n_{e}$ obtained from N and H$_{\alpha}$. This is much slower than expected from electron-ion recombination, which indicates that there is a significant production of electrons in the recombination phase. Additionally the decay of the N emission is significant longer than the lifetime of the excited states which means that highly excited N atoms are produced after the discharge. Both effects illustrate the importance of the energy stored in metastable and vibrational excited molecular and atomic nitrogen species. It would be very interesting to model this discharge. However, since a lot of initial densities (e.g.\ N$_{2}$(a$^{\prime}$) and excited N, N$^{+}$, N$_{2}^{+}$) are not known, the model will not give more information than the current estimates in this paper.

\ack The authors acknowledge the funding by STW (Stichting Technologische Wetenschappen). Loek Baede and Huib Schouten are acknowledged for their technical assistance and Manuel González for providing some simulated H$_{\alpha}$ Stark profiles with combined electron density and static electric field broadening effects.

\section*{References}

\bibliographystyle{iopart-num}
\bibliography{Bibliography}

\end{document}